\documentclass[reqno]{amsart}
\usepackage{graphics}
\usepackage[utf8]{inputenc}
\usepackage{amsmath}
\usepackage{pifont}

\newcommand{\be}{\begin{equation}}
\newcommand{\ee}[1]{\label{#1} \end{equation}}
\newcommand{\ba}{\begin{eqnarray}}
\newcommand{\ea}[1]{\label{#1} \end{eqnarray}}
\newcommand{\nl}{\nonumber \\}
\newcommand{\pd}[2]{ \frac{\partial {#1}}{\partial {#2}}  }
\newcommand{\re}[1]{(\ref{#1})}
\newcommand{\ol}{}

\usepackage{color}

\begin{document}

\title{First order and stable relativistic dissipative hydrodynamics}
\author{P. V\'an$^{1,2}$ and T.S. Bir\'o$^1$}
\address{$^1$Dept. of Theoretical Physics, KFKI Research Institute for Particle and Nuclear Physics, \\  H-1525 Budapest, P.O.Box 49, Hungary; 
and  {$^2$Dept. of Energy Engineering, Budapest Univ. of Technology and Economics},\\
  H-1111, Budapest, Bertalan Lajos u. 4-6,  Hungary}

\begin{abstract}
Relativistic thermodynamics is derived from kinetic equilibrium in a general frame. Based on a novel interpretation of Lagrange multipliers in the equilibrium state we obtain a generic stable but first order relativistic dissipative hydrodynamics. Although this was believed to be impossible, we circumvent this difficulty by a specific handling of the heat flow.
\end{abstract} 
\date{\today}
\maketitle 
\section{Introduction}\label{intro}

Thermodynamics, hydrodynamics and
kinetic theory are intimately related to each other. One may derive hydrodynamics from kinetic theory (via the Chapmann-Enskog expansion, or by momentum methods), but a certain interpretation  of the thermodynamic quantities and relations is required prior the derivation. One may obtain the Gibbs relation by a proper averaging in the framework of a dynamic theory for continuous media. At the same time, a definite concept of local equilibrium is necessary to calculate the entropy production and by doing to separate the dissipative and non-dissipative parts in the description. In this paper we explore these mutual relationships while  generalizing relativistic thermodynamics.
By a novel treatment of Lagrange multiplicators, defining the local equilibrium state, we derive a first order and still generic stable, relativistic dissipative hydrodynamics. 

Relativistic dissipative hydrodynamics is not a straightforward generalization of the corresponding nonrelativistic theory: the simplest relativistic analogue of the Fourier-Navier-Stokes system of equations   \cite{Eck40a3} is to date considered as conceptually wrong due to two reasons: it is unstable and acausal \cite{HisLin85a}. Thereafter generalized theories has been constructed by different methods, like   gradient expansion in the Landau-Lifshitz frame \cite{BaiEta06a,Rom10a,Rom10a1}, GENERIC \cite{Ott98a,Ott05b}, with the help of kinetic theory 
\cite{IsrSte76a,IsrSte79a,IsrSte79a1,MulRug98b,LiuAta86a,BetEta09a,BetEta11a}, by phenomenological extensions \cite{Isr76a,PavAta82a,JouAta92b} or  directly constructing divergence type equations \cite{PerCal10a}.  The  best known generalization, the Israel-Stewart theory \cite{Isr76a,IsrSte76a,IsrSte79a,IsrSte79a1}, extends the local equilibrium of Eckart theory improving its stability and causality properties \cite{HisLin83a,HisLin88a}.  In spite of  the motivation of the  extension, there is little known about the stability and causality of other extended theories with the famous exception of the divergence type theories \cite{GerLin90a,Ger95a,Ger01m}. These are namely causal by construction. 
In all the previously mentioned  cases, the the Fourier-Navier-Stokes system is the base, which is extended relativistically by the Eckart theory,  and extended further by the different methods listed above.

From the thermodynamic  point of view, the stability of the thermodynamic equilibrium of the fluid is indispensable: in the absence of external forces dissipation  always damps  perturbations.  This property is called generic stability. However, according to the investigations of Hiscock and Lindblom \cite{HisLin85a,His86a} the Eckart theory is unstable in this sense,  while the Israel-Stewart theory is conditionally stable. The appropriate conditions are far from being pure thermodynamical, their mathematical form is complicated and lack a simple physical interpretation. The nonrelativistic Fourier-Navier-Stokes equations are generic stable, and the conditions of stability are straightforward: concave entropy function and the nonnegativity of the transport coefficients.  In nonrelativistic hydrodynamics the first order theory is generic stable, while no first order generic stable theory was known in the relativistic case. This diametric contradiction in the stability properties between  nonrelativistic and relativistic fluids and the complexity of stability conditions of the extended theories indicate that the thermodynamic background is insufficiently understood.  

The  apparent departure between nonrelativistic and relativistic theories initiated several investigations regarding the possible reasons, not only to extend, but also to improve the Eckart theory. Garc\'ia-Col\'in and Sandoval-Villalbazo suggested to introduce a balance equation of internal energy in addition to the balance of energy-momentum \cite{GarSan06a,GarEta09a}. Muschik and Borzeszkowski noted, that this suggestion is not compatible with the foundations of relativity. We share their view, the energy-momentum density tensor has to include all energies of the system \cite{MusBor07a,GarSan07a}. Tsumura, Kunihiro and Onoshi \cite{TsuAta06a,TsuKun08a,TsuKun11a} used the renormalization  group method in order to reduce the relativistic Boltzmann equation for describing slow macroscopic dynamics. They concluded with the generic stability of their first order relativistic dissipative hydrodynamics in case of rarefied ideal gases in the Eckart frame. The obtained  set of of dissipative hydrodynamic equations is similar  for Garc\'ia-Col\'in et. al. and for Tsumura et. al. in the sense that heat flow is proportional to the gradient of the temperature like in the nonrelativistic Fourier's law, the acceleration related term of the Eckart theory is missing. In our approach this infamous term is not left out, but other terms compensate their destabilizing effect. The conditions  for  stability by missing accelerating term require the inequality \(\left.{\partial e/ \partial \mu}\right|_T> 0 \), which may be   valid for uniform ideal gases, but can be violated    in general, e.g. for Van der Waals gases.

\section{Eckart theory}
 
  Here and in the followings we use the metric $diag(1,-1,-1,-1),$ the units are chosen  such that the speed of light and the Boltzmann constant are unity $c=1$, $k_B=1$. In Eckart theory the thermodynamic description of the local equilibrium is simple and reasonable: one introduces the entropy density as a function of the local rest frame energy density \(e= u_aT^{ab}u_b\) and of the particle number density  \(n=u_a N^a\). We note that in relativistic particle systems  one of several conserved currents replace the particle number four-vector \(N^{a}\). The traditional continuity equation rather expresses the conservation of energy to  leading  order for massive (atomic) matter, \(e_0= mc^2 n\) being the rest frame energy. Thereafter \(u^{a}\) denotes the four velocity field actually used and \(T^{ab} \) is the energy-momentum density tensor.
The  Gibbs relation in the Eckart theory is
given as
\begin{equation}
de= Tds+\mu dn,
\label{EGrel}\end{equation}
where \(T\) is the temperature, \(s= u_a S^a\) is the entropy density, \(\mu\) is the chemical potential and \(S^{a}\) denotes the entropy four-current. 

Analyzing the structure of the second law inequality, we recognized that the entropy must be at least a function of the energy-momentum density vector \(E^{a}= T^{ab}u_b\),  instead of the energy density \(e\) \cite{Van08a,VanBir08a}. The particular case when the entropy depends on the Lorentzian length of the energy-momentum density vector, \(\sqrt{E^aE_a} = \sqrt{e^2+q_{a}q^a }\) leads to the following form of the Gibbs relation:
\be
 de + \frac{q_a}{e} dq^a = Tds+\mu dn, 
\ee{VBGrel}
where \(q_a = u_cT^{cb}\Delta_{ba}\) is the momentum density, the time-spacelike part of the energy-momentum density tensor with respect to  the local velocity field, and \(\Delta_{ba} = \delta_{ba}-u_au_b\).

The analysis of the structure of generic instabilities is also instructive in this approach. The energy-momentum density tensor of a single component fluid in a rest frame can be written as

\begin{equation}
T^{ab} = \begin{pmatrix} u_aT^{ab}u_b & u_cT^{cb}\Delta_{b}^a \\
\Delta^{a}_{c} T^{cb}u_{b} & \Delta^a_{c} T^{cd}\Delta_{d}^b \\
\end{pmatrix} = 
\begin{pmatrix}e & q^b \\
q^a & P^{ab} \\
\end{pmatrix}, 
\end{equation} 
where the space-spacelike part \(P^{ab}\) is the pressure, the current density of the momentum. For one component, neutral  fluids with zero spin, the energy-momentum tensor is  symmetric. As a consequence, the  time-spacelike and space-timelike components are equal\footnote{In  \(c=1\) system of units.}. However, the physical role of these spacelike vectors, \(q^{a}\) and \(q^{b}\), is different. One of them is a momentum density, the other is an energy current density in a given frame. In nonrelativistic theories only the second one is related to dissipation, the pure momentum density is not.  One may observe, that the instable modes of the Eckart theory are related to the momentum density. It is well known that the nonrelativistic approximation breaks the symmetry of these terms, momentum density and energy current density have different magnitudes when the speeds are low compared to the speed of light. However, the mentioned instabilities indicate a different hierarchy between the relativistically identical terms. The particular form of the local equilibrium expressed by (\ref{VBGrel}) breaks the symmetrical role of the momentum density and energy current density in dissipative relativistic fluids. We note that Eq. (\ref{VBGrel}) already improves  the stability properties of the corresponding  a relativistic dissipative fluid mechanics, the theory proves to be generic stable not only in the Eckart frame but also in case of general flows. Moreover, the generic stability does not require conditions beyond natural thermodynamic requirements: the thermodynamic stability and nonnegative transport coefficients \cite{VanBir08a,Van09a}. 

A third, independent justification for a generalized Gibbs relation like \re{VBGrel} above arises investigating  pure thermodynamic aspects. Analyzing the thermal interaction among  two thermodynamic bodies, one gains a physical interpretation of the different transformation properties of the temperature suggested and discussed by Planck and Einstein, Blanusa and Ott, Landsberg \cite{Pla07a,Ein07a,Bla47a,Ott63a,Lan66a,Lan67a},  later incorporating also the Doppler effect \cite{CosMat95a,LanMat96a,LanMat04a,BirVan10a}. Since in our approach the Gibbs relation is modified at the density level, our explanation is more natural than those propositions where the transformation properties would depend on the definition of the relativistic thermodynamic body \cite{Yue70a,CubAta07a,DunHan09a,DunEta09a}.

So far second law compatibility, generic stability of the fluid theory and the reasonable thermodynamic structure of the local equilibrium are only necessary conditions to obtain a proper fluid theory. A further aspect arises from the compatibility with the kinetic theory of gases.  In the following we explore from the point of view of kinetic theory, whether the generalized relativistic Gibbs  relation \re{VBGrel} can be further improved. Elaborating such an improvement  we derive the corresponding relativistic dissipative hydrodynamic theory, which preserves all the good stability properties of the previous generalization always in the first order approach.

\section{Thermodynamic equilibrium in kinetic theory}
\label{sec:2}

For the sake of simplicity, we consider in this paper a single-component Boltzmann gas without external fields. The mass of the particles is \(m\), their four-momenta are denoted by $p^a$,  $\sqrt{p^ap_a}=m$. The Boltzmann equation for the one particle  distribution function $f(x^a,p^a)$ is  given in the classical form
\be
 p^a\partial_a f = C(f).
\ee{Bo_eq}

Here $\partial_a$ denotes the partial derivative of $f$ regarding the spacetime variable $x^a$ and $C(f)$ is the following collision integral:\ 
\begin{equation}
C(x,p)= 
 \frac{1}{2} \int d\omega_1 d\omega' d\omega'_1 \left[
        f'f_1'W(p',p_1'|p,p_1)-ff_1W(p,p_1|p',p_1')\right], 
\label{coll_int}
\end{equation}
where $d\omega = \frac{d^3p}{p^0}$ is the momentum space measure.  $W(p,p_1|p',p_1')$ denotes the transition rate from  a  particle pair  momentum $(p', p'_1)$ before the collision, to particle pair momentum $(p,p_1)$ after the collision \cite{GroAta80b}. 

The particle number density four-vector, the energy-momentum density tensor and the entropy density four-vector are defined as 
\ba
N^a &:=& \int d\omega p^a f, \label{pnum_def}\\
T^{ab} &:=& \int d\omega p^ap^b f, \label{emom_def}\\
S^a &:=& -\int d\omega p^a f(\ln f-1). 
\ea{entr_def}

The local thermodynamic equilibrium is characterized by the condition of conserved entropy (\cite{GroAta80b}, page 43). Then it follows, that  the logarithm of the equilibrium distribution function is a collision invariant, therefore 
\be
{f}_0(x^a,p^b) = e^{\alpha-\beta_cp^c}
\ee{edistr_def}
with arbitrary fields  \(\alpha(x^a)\) and \(\beta^c(x^a). \)

For the equilibrium distribution function the particle number  balance gives: \be
 \partial_a N_0^a = \int d\omega p^a \partial_a f_0 =
  N_0^a\partial_a\alpha -T_0^{ab} \partial_a \beta_b =0.
\ee{pnbal_eq} 
In order to obtain thermodynamic relations, one can  calculate the equilibrium value of the Boltzmann-Gibbs entropy (\ref{entr_def}):    
\be
 S^a_0 =  \int d\omega p^a f_{0}(\ln f_{0}-1)=(1-\alpha)N_0^a +\beta_bT_{0}^{ab}.
\ee{Entreq}
 \re{pnbal_eq} and \re{Entreq} result in 
\be
\partial_aS_0^a + \alpha \partial_aN^a_0 - \beta_b\partial_aT_{0}^{ab} = 0.
\ee{en_eq}
The formulas \re{pnbal_eq}  and \re{en_eq} resemble a Legendre transformation, defining  this way the thermodynamic structure of the theory with the five Lagrange multipliers \(\alpha\) and \(\beta_b\) \cite{Isr63a}.  However,   \re{en_eq} cannot be interpreted as a generalized Gibbs relation. This relation alone does not fix the entropy as a function of energy-momentum and particle number density. To find thermodynamic relations, one introduces a flow, a velocity field over the continuum material.

According to the usual interpretation of the equilibrium distribution function (\ref{edistr_def}), such a vector field appears through the vectorial invariant \(\beta_a\),  choosing that the velocity field parallel to it. However,  in the general case, \(\beta_a\) decomposes into components parallel and perpendicular to the velocity field
\be
\beta^a = \frac{u^a +w^a}{T}.
\ee{decbeta}

There we have used the notation \(1/{T}:= \beta_au^a\) and \(w^a := {\Delta^a_b \beta^b}/{\beta_cu^c}\). It is convenient to introduce \(g^a = u^a+w^a\) as well. We can see that 
$g^a u_a = 1$  and \(w^{a}u_a=0\), moreover, $w_aw^a\leq1$, because \(\beta^a\), and therefore \(g^a\), are timelike. With \(\mu = \alpha T\),  the  equilibrium distribution can be written as   
\be
f_0 = e^{\frac{\mu-p^a(u_a+w_a)}{T}}. 
\ee{genJut_def}
 
The equilibrium particle number density four-vector and energy-momentum density tensor can be calculated recognizing that the usual velocity field, defined by the direction of \(\beta ^{a}\), is 
$$
\hat u^a =\frac{\beta^a}{\sqrt{\beta^a\beta_a}}=\frac{u^a+w^a}{\sqrt{1-w^2}},
$$ 
where \(w^2= -w^aw_a\). In the following this usual equilibrium velocity field \(\hat u^a\) is referred to as the \textit{natural frame}.

 In the general frame, the equilibrium particle number four-vector $N^a_0$ and energy-momentum density tensor \(T_0^{ab}\)  are
given as\ba
N_0^a &=& ng^{a} = n u^{a} + n w^a, \label{ndpnumg}\\
T_0^{ab} &= & (e+p)g^ag^b - p \delta^{ab} \nonumber\\
    &=& e u^{a} u^b + q^{a}u^b+q^bu^a - p\Delta^{ab}+\frac{q^aq^b}{e+p}.
\ea{ndemomg}
Here, the thermodynamic quantities in the general frame (without hat) are related to the usual quantities in the natural frame (with hat) as \( T=\sqrt{1-w^2} \hat T\), \(\  \mu=\sqrt{1-w^2} \hat \mu\), \(\ n ={\hat n}/{\sqrt{1-w^2}}\), \(\ p = \hat p\), \(\ e = {(\hat e+\hat p w^2)}/{(1-w^2)}\), (see \cite{Van11p} for more details). The relation between the momentum density and the particle number current density is fixed:
\be
q^a = (e+p)w^a.
\ee{heatdiff}

Therefore, for nondissipative fluids, the Eckart frame and the Landau-Lifshitz frame coincide. Substituting \re{ndpnumg} and \re{ndemomg} into \re{Entreq} 
yields  
\be
\! S^a_{0} =\frac{g^a}{T} \left ( nT -\mu n + E^bg_b\right)  ,
\ee{sid2}
where \( E^a = eu^a + q^a\). The entropy current density in the general frame is parallel to \(g^{a}\).  The thermodynamic relations are calculated by requiring the following inequality \cite{Isr63a,Van11p}: substituting \re{ndpnumg} and \re{ndemomg} into \re{en_eq}:
\begin{multline}
\partial_aS_0^a + \alpha \partial_aN^a_0 - \beta_b\partial_aT_{0}^{ab} = \\
 \frac{g^a}{T}  \left(T\partial_a  s+ \mu\partial_a n - 
    g_b \partial_a E^b - pw^b \partial_au_b\right)+ \frac{\partial_a g^a}{T} \left(T s +\mu n -g_bE^b- p \right)=0. \label{devbasic2}
\end{multline}

Now the first term of the right hand side of the previous equality does not contain contracted derivatives, and allows to extract thermodynamic relations. One recognizes that the entropy density in the general frame is the function of the particle number density, of the energy-momentum density four-vector and of the velocity  \(s =  s(E^a,  n, u^a)\), with the following partial derivatives:
\be
\pd{s}{n} = \alpha=\frac{\mu}{T}, \qquad
\pd{s}{E^a} = \frac{g_a}{T}, \qquad
\pd{s}{u^a} = \frac{pw_a}{T}.
\ee{classpd1}
The consequent Gibbs relation for the differentials is
\be
 g_{a}d E^{a} +p w_{a} du^{a}= de+w_a dq^a +[(e+p)w_a -q_a]du^a=Tds + \mu d n .
\ee{Gr_res}
 
Utilizing Eq. (\ref{heatdiff}), the Gibbs relation defining the local equilibrium of a relativistic fluid is given by 
\be
de +\frac{q_a}{h}dq^a = Tds +\mu dn, 
\ee{Grel}
where \(h=e+p\) is the enthalpy density. Whenever  \(w^{a}\) is zero, we obtain  (\ref{EGrel}) as a special case. This unusual Gibbs-relation  already considers both energy-momentum exchange in the form of transport and also in the form of mechanical work. The Legendre transformation properties are fixed by the last bracket in \re{devbasic2}. It vanishes if
\be
  p +g_aE^a= p+e + \frac{q^{a}q_a }{h}=  Ts +\mu  n . 
\ee{potrelg}

 A comparison to  \re{sid2} provides the thermal equation of state of our ideal gas: \( p =  nT\). This equation of state is valid in general frames.

\section{Dissipative hydrodynamics}

In order to determine the constitutive functions of the hydrodynamic theory in a linear Onsagerian framework, we calculate the entropy production with the help of the generalized Gibbs relation (\ref{Grel}).  Introducing the substantial or comoving time derivative $\frac{d}{d\tau}:= u^a\partial_a$, denoted by an overdot,  the basic balance equations are 
\ba
\partial_a N^a &=& \dot n + n\partial_au^a + \partial_aj^a =0,\label{pnumc_bal}\\
\partial_b T^{ab} &=&  \dot e u^a +eu^a\partial_bu^b +\dot q^a +q^a\partial_b u^b +
             e\dot u^a   \nl
        &&     +\ u^a\partial_b q^b+q^b\partial_b u^a +\partial_bP^{ab}= 0^a. 
\ea{emomc_bal}

The energy and three-momentum balances are the time- and spacelike components of the energy-momentum balance: 
\ba
u_a\partial_b T^{ab} &=&  \dot e  +e\partial_bu^b +u_a\dot q^a +\partial_b q^b +u_a\partial_bP^{ab} = 0, \label{e_bal}\\
\Delta^a_c\partial_b T^{cb} &=&   e\dot u^a +\Delta^a_b\dot q^b +q^a\partial_b u^b + q^b\partial_b u^a +\Delta^a_c\partial_bP^{cb} =0^a. 
\ea{momc_bal}
The  calculation of the entropy production is now straightforward. The time derivative of the entropy density is determined by using the  functional dependencies inherent in the Gibbs relation (\ref{Grel}). Then we substitute the  balance equations (\ref{pnumc_bal}) and (\ref{e_bal}) and we obtain 
\begin{gather}
\partial_a S^{a} =  \dot s  +s\partial_bu^b +\partial_a J^a 
       +\partial_a\left(\frac{q^a-\mu j^a}{T}\right)  \nonumber\\
    =\frac{1}{T}\left(P^{ab}-\left(e+\frac{q_cq^c}{h}-Ts-\mu n\right)\Delta^{ab}\right)\partial_au_b         +q^a\left(\partial_a\frac{1}{T} +\frac{\dot u_a}{T} +\frac{\dot q_a+q_a\partial_bu^b}{hT}\right)\nonumber\\
    = -j^a\partial_a\frac{\mu}{T} + \frac{1}{T}\Pi^{ab}\partial_au_b+
        q^aX_{th}^a\geq 0.
\label{entrc_bal}\end{gather}
Here the static pressure was determined by \re{potrelg},  \(\Pi^{ab}=P^{ab}+p\Delta^{ab}\) is the viscous stress and  
\be
X_{\rm th}^b = \Delta^{ba}\left(\partial_a\frac{1}{T} +
    \frac{1}{hT}\left(h\dot u_a+\dot q_a + q_a\partial_c u^c \right)\right)
\ee{thforce}
is the  thermodynamic force of the thermal interaction. The classical form of the entropy current density $J^a = \frac{1}{T}(q^a -\mu j^a)$ in \re{entrc_bal} can be justified by a deeper analysis of the second law \cite{Van08a}.

Consequently assuming linear relationship
between the thermodynamic fluxes and forces leads to the following constitutive relations:
\be
  q^a = \lambda X^{a}_{\rm th},\quad  
  j^a =-\xi \Delta^{ab}\partial_b\frac{\mu}{T},\quad
  \Pi^{ab} = -2\eta \partial^{\langle a} u^{b\rangle}-\zeta\partial_c u^c\Delta^{ab},
\ee{linOn}
where \(\lambda\) is the heat conduction coefficient, \(\xi\) is the diffusion coefficient, \(\eta, \zeta\) are the shear and bulk viscosities, respectively. The bracket denotes the symmetric traceless part of the spacelike tensor
$$
\partial^{\langle a} u^{b\rangle}=
        \Delta^a_c\Delta^b_d\left(\frac{\partial^c u^d+\partial^d u^c}{2}-
        \frac{1}{3} \partial_eu^e \Delta^{cd}\right).
$$
The remarkable part of the entropy production formula is \re{thforce}. The gradient of the inverse temperature and the acceleration solely constitute the driving force of thermal interaction in the Eckart theory. The   additional terms containing the momentum density \(q^{a}\) are absent there. The terms beyond the gradient of the temperature stem from the first part of the momentum balance (\ref{momc_bal}) divided by the enthalpy density. This indicates that by relativistic heat conduction  the momentum and energy transport cannot be separated. 

\section{Generic stability}

A necessary, expected requirement of fluid theories is generic stability, that is the linear stability of the homogeneous equilibrium of the  system of equations  \re{pnumc_bal}, \re{e_bal}, \re{momc_bal} and (\ref{linOn}).
In the following we sketch the proof of this property in the Eckart frame, where \(j^{a}=0^a\).

Denoting the perturbed fields by $\delta$ we can put the linearized equations \re{pnumc_bal}, \re{e_bal}, \re{momc_bal} and the first and last equations of \re{linOn} into the following form
\ba
 0 &=& \dot{\delta n} +
        \ol n \partial_a \delta u^a , \label{stab_lnbal}\\
 0 &=& \dot{\delta e} +
        \ol h \partial_a \delta u^a +
        \partial_a \delta q^a, \label{lteb}\\
 0 &=& {\ol h}\dot{\delta u^a}  +
        \Delta^{ab} \partial_b \delta p +
        \dot{\delta q^a} +
        \Delta^a_c \partial_b \delta\Pi^{cb},
        \label{ltib}\\
 0 &=& \delta q^a-
        \lambda \Delta^{ab}\left(\partial_b \delta\frac{1}{T} -
         \frac{h\dot{\delta u}_b+\dot{\delta q}_b}{T} \right), \label{lfum}\\
 0 &=&  \delta\Pi^{ab} +
       {\eta}_v \partial_c \delta u^c \Delta^{ab} +
        \eta \Delta^{ac}\Delta^{bd}(
            \partial_c\delta u_d +\partial_d\delta u_c-
            \frac{2}{3}\partial_f\delta u^f\Delta_{cd}). 
\ea{lnewm}
 Now we are looking for solutions in the form of exponential plane wave form $\delta Q =  Q_0 e^{\Gamma t + i k x}$. Then the condition of asymptotic stability of the above equations is that the solutions of the characteristic equation \( \det {\bf M} = (\det {\bf N})(\det {\bf R})^2=0\) give negative real parts for \(\Gamma\). Here 
\be
  {\bf R} = \begin{pmatrix}
  h\Gamma          & \Gamma                       & ik    & 0\\
  \tilde\lambda T \Gamma      & 1+ \Gamma\frac{T \tilde\lambda}{h}  & 0     & 0\\
  ik\eta            & 0                         & 1     & 0\\
  ik\tilde{\eta}_v  & 0                         & 0     & 1\\
  \end{pmatrix},
\ee{R}
couples the perturbation fields \(\delta u^a, \delta q^a, \delta \Pi^{xa},\delta \Pi^{aa}\), \(a=y,z\) and 
\be
  {\bf N} = \begin{pmatrix}
  \Gamma & 0   & ikn      & 0    & 0 \\
  0   & \Gamma & ikh  & i k  & 0  \\
  ik\partial_n p        & ik\partial_e p  & \Gamma h & \Gamma  & ik \\
  ik\tilde\lambda_{}\partial_n T & ik\tilde\lambda\partial_e T
        &  \Gamma \tilde\lambda T     & 1+\frac{\tilde\lambda T}{h}\Gamma & 0 \\
  0   & 0   & ik\tilde{\eta}  & 0 & 1 
   \end{pmatrix}.
\ee{N}
couples \(\delta n, \delta e, \delta u^x, \delta q^x, \delta \Pi^{xx}\). Here \(h=e+p\),  $\tilde{\eta} = \eta_v +\frac{4}{3}\eta$,  $\tilde{\eta}_v = \eta_v -\frac{2}{3}\eta$ and \(\tilde \lambda = \lambda/T^2\). The determinant of \textbf{R} gives the condition $\left(h+ k^2 \frac{\tilde{\eta}\tilde \lambda T}{h}\right)\Gamma +
 \tilde{\eta} k^2 = 0$, which results in negative real \(\Gamma\) solutions for all wave numbers \(k\).
The determinant of {\bf N} results in the following condition
\begin{gather*}
 \Gamma^3 \left(h + k^2\frac{\tilde\lambda T}{h}\tilde{\eta} \right) + 
 \Gamma^2k^2 \left(\tilde{\eta} -
        \tilde\lambda  n \partial_nT+
        \frac{\tilde\lambda T }{h} n \partial_n p  \right) +\\
 \Gamma k^2 \left(h\partial_e p + n \partial_n p  +
        k^2 \tilde{\eta}\tilde\lambda\partial_e T \right)+  k^4\lambda n \left(  
        \partial_n p\partial_eT-\partial_e p \partial_n T\right)= 0.
 \end{gather*}

According to Routh-Hurwitz criteria the real parts of the roots of the polinomial equation $a_0 x^3+a_1 x^2 + a_2 x + a_3 =0$ are non negative if the coefficients are all non-negative and $ a_1 a_2 - a_0 a_3> 0$ \cite{KorKor00b}. It is straightforward to check  that these inequalities are fulfilled, if the transport coefficients \(\lambda, \eta, \eta_v\)are nonnegative and the inequalities of the thermodynamic stability are satisfied, i.e. $\partial_e T > 0$, $\partial_n \frac{\mu}{T} > 0$ and also $\partial_e T \partial_n \frac{\mu}{T} -
\left(\frac{\partial_nT}{T}\right)^2 \geq 0$. 

One may prove similarly  the generic stability of the above hydrodynamic equations in case of general frame, too.

\section{Summary and conclusions}

In nonrelativistic mechanics three-momentum and three-velocity are parallel, in relativistic mechanics the corresponding respective four-vectors are orthogonal. While in nonrelativistic mechanics the momentum is always zero in a local rest frame -- from \(\bf v=0\) follows \(m\bf v=0\) --, in a relativistic theory it is not necessarily so, there is only a particular rest frame where the momentum vanishes.  Therefore, in case of thermal interactions between at least two participants, one of them will see a nonzero exchanged momentum.   That is the basic reason why relativistic and nonrelativistic thermodynamics are different, and that is what a proper relativistic concept of local equilibrium should reflect.  It is still a local equilibrium by definition, since the entropy production is zero, and the generalized Gibbs relation requires a constant velocity of the thermodynamic body. However, it depends on the rest frame momentum because the relativistic concept of local equilibrium should be formulated in a general frame.  
 
In this paper we have shown that a particular extension of the concept of local equilibrium is compatible with kinetic theory when considering general flows. A corresponding Gibbs relation and the related thermodynamic formalism introduced the rest frame momentum density \(q^{a}\) as an additional density variable and the related intensive quantity \(w_{a} = q_a/h\) is a particular spacelike quantity with the physical dimension of velocity. The necessity of the distinction between momentum density and energy current density in a dissipative theory was already recognized in \cite{VanBir08a} and Eq. (\ref{VBGrel}) was suggested as the basic formula defining the local equilibrium. Our present study of relations to kinetic theory revealed that in local equilibrium inertia  is defined through the balance of momentum, therefore, the enthalpy density substituts the energy density in the properly generalized Gibbs relation (\ref{Grel}). 

From the point of view of the traditional concept of local equilibrium by Eckart, our approach might be considered as second order in the heat flow. According to the definition given by Hiscock and Lindblom \cite{HisLin85a} the obtained relativistic dissipative hydrodynamics is first order, because the deviations from the local equilibrium are first order in the entropy current. 

The generic stability of the equations is proved in the Eckart frame.

\section{Acknowledgement}   
The work was supported by the grants Otka K68108, K81161 and TÉT 10-1-2011-0061. The authors thank T. Fülöp, E. Molnár, W. Muschik, H.H.~v. Borzesz\-kowski and A. Muronga for valuable discussions.

\bibliographystyle{unsrt}

\end{document}